\documentclass[reprint,aps,prd,showpacs,preprintnumbers,amsmath,amssymb]{revtex4-1}
\usepackage{graphicx}
\usepackage{graphics}
\usepackage{subfigure}
\usepackage{slashbox}
\usepackage{amsmath}
\usepackage{brackets}
\usepackage{psfrag}
\usepackage{epsfig}
\usepackage{lineno}
\usepackage{hyperref}
\usepackage{amssymb}
\begin{document}

\ProvideTextCommandDefault{\textonehalf}{${}^1\!/\!{}_2\ $}

\title{Search for an axion-induced oscillating electric dipole moment for electrons using atomic magnetometers}

\author{P.-H.~Chu}
\email[Email address: ]{pchu@lanl.gov}
\author{Y.~J.~Kim}
\email[Email address: ]{youngjin@lanl.gov}
\author{I.~Savukov}
\affiliation{Los Alamos National Laboratory, Los Alamos, New Mexico 87545, USA}
\date{\today}

\begin{abstract}
We propose an experimental search for an axion-induced oscillating electric dipole moment (OEDM) for electrons using state-of-the-art alkali vapor-cell atomic magnetometers. The axion is a hypothesized new fundamental particle which can resolve the strong charge-parity problem and be a prominent dark matter candidate. This experiment utilizes an atomic magnetometer as both a source of optically polarized electron spins and a magnetic-field sensor. The interaction of the axion field, oscillating at a frequency equal to the axion mass, with an electron spin induces a sizable OEDM of the electron at the same frequency as the axion field. When the alkali vapor is subjected to an electric field and a magnetic field, the electron OEDM interacts with the electric field, resulting in an electron spin precession at the spin's Larmor frequency in the magnetic field. The resulting precession signal can be sensitively detected with a probe laser beam of the atomic magnetometer. We estimate that the experiment is sensitive to the axion-photon interaction in ultralight axion masses from $10^{-15}$ to $10^{-10}$~eV. It is able to improve the current experimental limit up to 5 orders of magnitude, exploring new axion parameter spaces.

\end{abstract}

\maketitle
\section{Introduction}
The non-zero electric dipole moment (EDM) of elementary particles is a direct evidence for violations of both parity ($P$) and time-reversal ($T$) symmetries~\cite{purcell:1950,landau:1957}; $T$ violation also implies charge-parity ($CP$) violation assuming the $CPT$ invariance~\cite{Lehnert:2016}. The demand of new $CP$-violation sources~\cite{Sakharov:1991} to explain the matter-antimatter asymmetry in the Universe~\cite{Canetti:2012} sparked the great interest in the EDM measurements. The EDMs of elementary particles have been searched for decades~\cite{Chupp:2017} since the first neutron EDM experiment in 1950~\cite{purcell:1950}. The extremely small upper limit of the neutron EDM~\cite{Baker:2006,pendlebury:2015} implied an unsolved strong $CP$ problem in the quantum chromodynamics (QCD)~\cite{peccei:1977}. The axion, a hypothetical elementary particle, was introduced to potentially resolve the strong $CP$ problem by Peccei and Quinn~\cite{peccei:19771,weinberg:1978}. 
 
The axion is a prominent candidate for dark matter in the Universe~\cite{Duffy:2009}. The evidence for dark matter came from cosmological and astrophysical observations, including the cosmic microwave background (CMB) power spectrum~\cite{Efstathiou:1992}, cluster and galactic rotation curves~\cite{zwicky:1933, rubin:1983}, gravitational lensing~\cite{walsh:1979,clowe:2006}, and large-scale structure formation~\cite{springel:2006}. The invisible dark matter has been known to compose more than 25\% of total mass-energy in the Universe~\cite{Aghanim:2018eyx}. Other possible candidates of dark matter include weakly interacting massive particles (WIMPs)~\cite{steigman:1985}, sterile neutrinos~\cite{kusenko:2009} and others~\cite{patrignani:2017}. Despite many experiments, the nature of the dark matter still remains unknown. 

The axion resolving the strong $CP$ problem is the so-called QCD axion. The mass of the QCD axion is given by 
\begin{align}
m_a \sim 6\times10^{-10}~\text{eV}\Big(\frac{10^{16}~\text{GeV}}{f_a}\Big)
\end{align}
where $f_a$ is the symmetry-breaking energy scale~\cite{peccei:1977}. The traditional range of axion mass is considered to be from $10^{-6}$ to $10^{-2}$~eV, the so-called ``axion window," according to constraints from observations of current experiments, astrophysics, and cosmology~\cite{Raffelt:2006}. The lower bound of the axion window is derived from cosmology dark matter abundance~\cite{Turner:1990}. However, theories beyond the Standard Model of particle physics, e.g., supersymmetry and super string~\cite{svrcek:2006,jaeckel:2010}, suggest that ultralight axions with mass much smaller than $10^{-10}$~eV can be generated by a broken global symmetry at high-energy scales such as the grand-unified ($f_a\sim10^{16}$~GeV) or the Planck scales ($f_a\sim10^{19}$~GeV). Besides, if the axion mechanism happened before the cosmic inflation, the inflation could have  highly suppressed the axion field and generate ultralight axions~\cite{Pi:1984,Linde:1991}. These axions, the so-called axion-like-particles (ALPs), do not have a specific relation between their field and mass, unlike the QCD axion. Axions, in principle, can couple with ordinary particles such as photons, electrons, and nucleons. Axions have been searched for decades in various systems; however, a large area of parameter space of the axion field amplitude and the axion mass has not been constrained, especially at the sub-eV range of axion mass. To this end, further sensitive experimental searches for axions are necessary. 

The axion field was first suggested as a classical field oscillating at a frequency equal to the axion mass $m_a$~\cite{preskill:1983,dine:1983} (the axion Compton frequency in natural units where $c=\hbar=1$), expressed as $a_0\cos(m_a t)$. Here $a_0$ is the local amplitude of the axion field. Recently, several papers have suggested that the interactions of the coherently oscillating axion field with particles of gluons and fermions will induce an EDM of the particles oscillating at the same frequency as the axion field~\cite{graham:2011,graham:2013,stadnik:2014,hill:2016,Alexander:2017}. The laboratory experiments detecting these oscillating effects will be sensitive to the axion mass~\cite{dine:1983,Duffy:2009}. Besides, while the effects of several other axion searches, such as axion helioscopes and light-shining-through-walls experiments, scale quadratically with the axion coupling constant~\cite{Graham:2015}, the effects of the oscillating axion field scale linearly with the axion coupling constant, as indicated in Eqs.~\ref{eq:oedm_nuclear} and~\ref{eq:oedm_electron} described below~\cite{budker:2014,abel:2017}. 

The cosmic axion spin precession experiment (CASPEr) has been proposed to search for a nuclear oscillating  EDM (OEDM) induced by axions~\cite{budker:2014}. This experiment is based on the nuclear magnetic resonance technique. The nuclear OEDM, $d_n$, is given by~\cite{graham:2013}
\begin{align}
    d_n =& g_d a(t) = g_d a_0 \cos(m_a t)
    \label{eq:oedm_nuclear}
\end{align}
where $g_d$ is the coupling strength of the axion-gluon interaction and $a(t)$ is the local axion field oscillating at the axion mass. On the assumption that this axion field composes all dark matter, its density will be equal to~\cite{graham:2013} 
\begin{align}
    \rho_{DM} = \frac{1}{2}m_a^2 a_0^2
\end{align}
where $\rho_{DM}\sim 0.3$ GeV/cm$^3$~\cite{patrignani:2017}. Then, the nuclear OEDM can be rewritten as 
\begin{align}
    d_n
    =& g_d\frac{\sqrt{2\rho_{DM}}}{m_a} \cos(m_a t).
    \label{eq:dn}
\end{align}
The CASPEr experiment utilizes nuclear spins in a solid sample, pre-polarized by an external strong magnetic field $B_0$. The experiment is conducted in a low temperature in order to reduce thermal noise. In the presence of an electric field, perpendicular to $B_0$, the polarized nuclear spins can be rotated away from the direction of $B_0$ due to the interaction of the nuclear OEDM with the electric field. It causes the spins to precess at the spin's Larmor frequency in $B_0$, which can be detected by a sensitive magnetometer such as a superconducting quantum interference device (SQUID)~\cite{budker:2014} or an atomic magnetometer (AM)~\cite{wang:2018}, both of which can reach a 1 fT/$\sqrt{\text{Hz}}$  sensitivity. 

Although the main focus of CASPEr is axion coupling to nucleon~\cite{budker:2014}, the axion could also couple to electrons~\cite{hill:2016} or other particles~\cite{stadnik:2014}.  According to Refs.~\cite{Hill:2015,hill:2016,Hill:2017}, the interaction between an electron spin and the axion field can induce a $CP$-violating, non-zero electron OEDM. A classical derivation in the Maxwell equations is described in Ref.~\cite{Hill:2016zos}. The axion-induced electron OEDM is written as
\begin{align}
    \vec{d_e}(t) =& g_{a\gamma\gamma} 2\mu_{\text{B}}a_0\cos(m_a t)\hat{\sigma}
\label{eq:oedm_electron}
\end{align}
where $g_{a\gamma\gamma}$ is the strength of the axion-photon coupling~\footnote{The electron OEDM is defined as $d_e(t) = 2g_a \Big(\frac{a_0}{f_a}\Big)\mu_{B} \cos(m_at)$ in~\cite{hill:2016}. The author also suggested that the electron OEDM could be three orders of magnitude larger than that of the nucleon owing to larger magnetic moment of electrons than nucleons. Here we define $g_{a\gamma\gamma} \equiv g_{a}/f_a$ in order to compare with other experiments. For the QCD axion in the KSVZ model ~\cite{patrignani:2017}, $g_{a\gamma\gamma}\sim 4\times 10^{-10} (\frac{m_a}{\text{eV}}) [\text{GeV}^{-1}]$.} and $\mu_\text{B}$ is the Bohr magneton. The OEDM is collinear with electron spin $\vec{\sigma}$. The amplitude of the OEDM can be estimated from Eq.~(3):
\begin{align}
    d_e=8.3\times10^{-23}~e\cdot\text{cm}\Big(\frac{g_{a\gamma\gamma}}{\text{GeV}^{-1}}\Big)\Big(\frac{\text{eV}}{m_a}\Big).
    \label{eq:de-Amp}
\end{align}
The coupling between axions and electrons, the $g_{aee}$ term, can also induce the spin precession without the need of an electric field as described in Eq.~21 of Ref.~\cite{graham:2013}. Based on the current constraint $g_{aee} < 10^{-10}~\text{GeV}^{-1}$, the effective energy shift due to the axion-electron coupling is about 5 orders of magnitude smaller than the sensitivity of the energy shift of the proposed experiment (see Eq.~9).  Therefore, we neglect the contribution from the axion-electron coupling. 

The static electron EDM has been searched in various systems such as atoms~\cite{Chin:2001,Regan:2002}, molecules~\cite{Hudson:2011,Baron:2014}, ions~\cite{Cairncross:2017}, and solid-state materials~\cite{Eckel:2012, Kim:2015}. So far the best constraint for electron EDM is $8.7\times 10^{-29}~ \text{e}\cdot\text{cm}$ using polar molecules~\cite{Baron:2014}. The rubidium (Rb) and potassium (K) AMs have been also suggested to employ for hunting the static electron EDM because of their high sensitivity~\cite{Weiss:2003}. While SQUIDs have been proposed as a magnetic-field sensor~\cite{Alexander:2018}, in this paper, we propose to search for the axion-induced electron OEDM using a new technique based on AMs, a non-cryogenic magnetic-field sensor. 

\section{Experimental Approach}
\begin{figure}[b]
\includegraphics[width=0.45\textwidth]{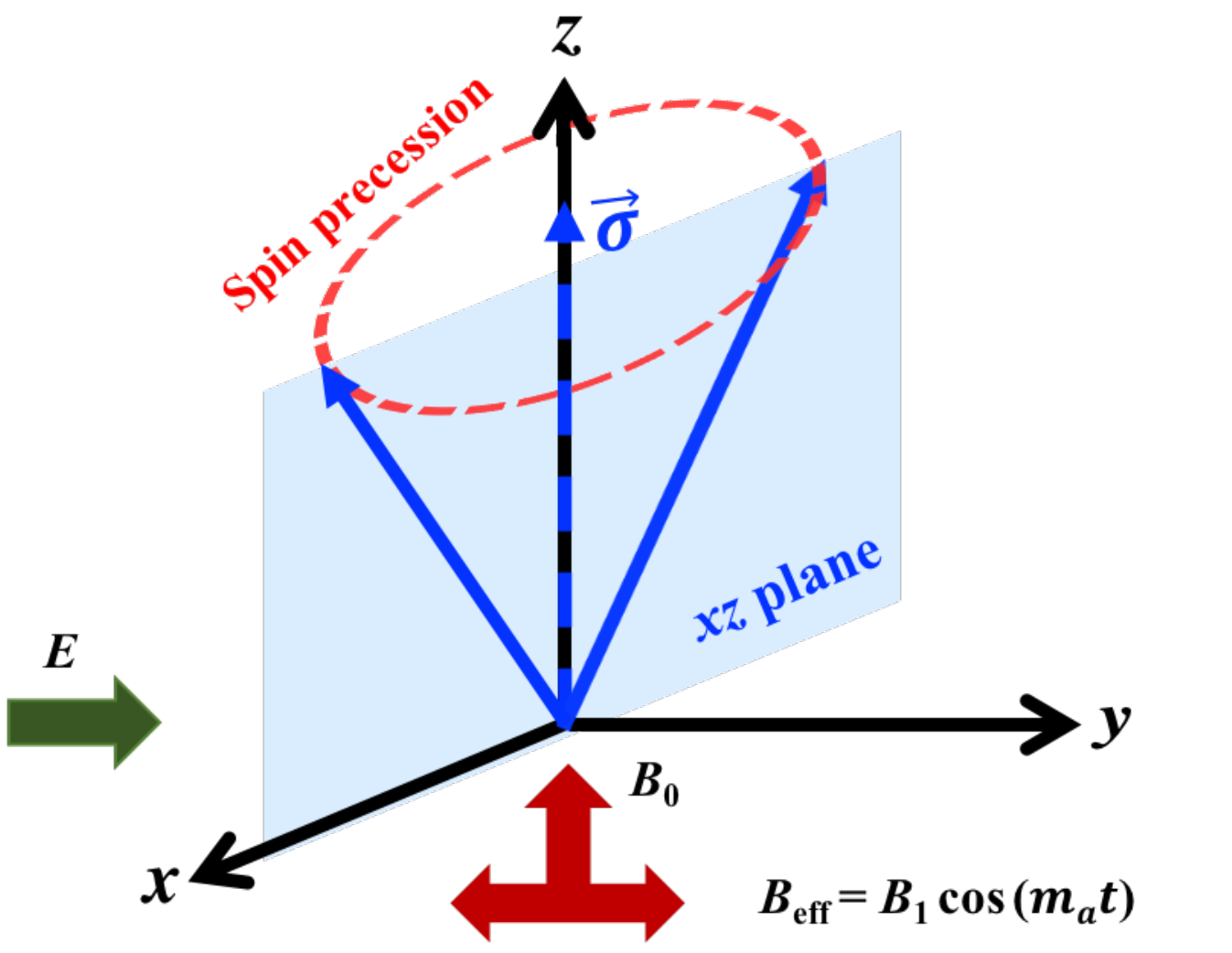}
\caption{Scheme of detecting an axion-induced electron OEDM using an AM with pumping in the $z$ direction and probing in the $x$ direction. A magnetic field and an electric field are externally applied to the AM vapor along the $z$ and $y$ direction, respectively. The interaction of the electron OEDM with the electric field generates an effective oscillating magnetic field, $\vec{B}_{\text{eff}}$, in the $y$ direction, which causes the electron spins to precess at the spin's Larmor frequency. The projection of the spin precession onto the $x$ direction is the response of the AM to the electron OEDM. }
\label{fig:bloch}
\end{figure}
The basic principle of the axion-induced electron OEDM search using an AM is illustrated in Fig.~\ref{fig:bloch}. One important advantage of this experiment is that an AM will serve as both a source of polarized electrons and a very sensitive detector, so it does not require an additional solid sample and a pickup coil system, for example, the one in the CASPEr experiment. Electron spins of an AM alkali vapor are polarized by a pump laser beam in the $z$ direction. To determine the Larmor frequency for the polarized electron spins, a holding magnetic field $\vec{B}_0$ is externally applied to the vapor along the pump beam $z$ direction. Unlike the CASPEr experiment, this experiment does not require a several Tesla superconducting magnet for polarization of electron spins, greatly simplifying the experimental design. 

In order to observe the effect of the electron OEDM, a static electric field $\vec{E}$ is externally applied to the vapor in the direction perpendicular to $\vec{B}_0$, in the $y$ direction. The electron OEDM interacts with the electric field,  $-\vec{d_e}(t)\cdot\vec{E}$, which induces an effective oscillating magnetic field
\begin{align}
    \vec{B}_{\text{eff}} =  [{E} g_{a\gamma\gamma} 2\mu_{\text{B} }a_0 \cos(m_a t)/\gamma]\hat{y}\equiv B_1 \cos(m_a t)\hat{y}
    \label{eq:beff}
\end{align}
where $\gamma$ is the gyromagnetic ratio of the AM alkali atoms. The total magnetic field at the AM vapor is written as $\vec{B}=B_0\hat{x}+B_1\cos(m_a t)\hat{y}$. The effective field $\vec{B}_{\text{eff}}$ causes the electron spins to tilt away from the direction of $\vec{B}_0$ and to precess. Such spin precession can be sensitively detected with a probe laser beam in the $x$ direction. When the spin's Larmor frequency is equal to the axion mass, the precession signal will be maximized because a resonance occurs. The goal of this experiment aims to precisely detect the axion-induced spin precession with an AM, which will determine or constrain the axion-photon interaction.

For application of an electric field to an AM vapor, two conductive planar electrodes will be placed on the vapor cell. To reduce the magnetic Johnson noise, electrodes will be made from machinable ceramics coated with graphite that has large electrical resistivity~\cite{Kim:2015}. The sensitivity of the OEDM experiment to the axion-photon interaction can be improved by increasing the strength of the electric field (see Eq.~\ref{eq:beff}). One electrode is connected to a high-voltage source and the other electrode to ground. To prevent electrical breakdown of the vapor cell, the cell can be filled with some pressures of nitrogen gas~\cite{Meek1953}. According to Ref.~\cite{Seltzer2008}, it was measured that the amplitude of an electric field inside a vapor cell is smaller than the applied field amplitude. This requires to precisely monitor the field amplitude during the OEDM experiment by, for example, measuring the Stark shift of the optical resonance frequency of a probe laser beam ~\cite{Seltzer2008}. In addition, it was found that a high electric field can adversely affect the alkali vapor density~\cite{Seltzer2008}; the vapor density in a cell made of uncoated aluminosilicate glass slowly decreased in the presence of the electric field in the cell, depending on the strength of the field, and the vapor density did not immediately recover when the field was removed. Possible solutions for the density decrease are to coat the interior cell wall with multilayer of octadecyltrichlorosilane or make the cell with Pyrex glass~\cite{Seltzer2008}. The electric field homogeneity in such glass vapor cells can be improved by a silane coating on the interior cell walls~\cite{Hunter1988}.  

The OEDM experiment eliminates common systematics in static EDM experiments, which are associated with spurious magnetic fields that mimics an EDM. The systematic fields are produced mainly due to electric-field reversals or movement of target particles~\cite{Kim:2015}. A possible systematic in the OEDM experiments will be transverse fluctuations of the holding field $\vec{B}_0$, which can be suppressed by using a low-noise voltage source, such as dc batteries, and by designing a better coil system such as Maxwell coils or multi-coil arrays~\cite{Wang2002,Merritt1983,Garrett1967}. The transverse component of the holding field produced by such coil systems can be reduced, suppressing its fluctuations. Another possible systematic will be mechanical vibrations at the target frequency, which can be eliminated by making a rigid experimental setup. The dc magnetic field generated from a leakage current across the vapor cell is the most important systematic in static EDM experiments, while in OEDM experiments, it can only change the resonance frequency of an AM, given by $f_0=\gamma \sqrt{B_x^2+B_y^2+B_z^2}$ where  $B_{x,y,z}$ are the dc magnetic field components in the $x$, $y$, and $z$ direction ~\cite{Karaulanov:2016}. For example, in an experiment using an AM with a 1~cm-diameter Rb cell, a 100~Hz bandwidth, and $\gamma$ of $7\times10^{9}$~Hz/T: if a leakage current flows in a loop around the cell, the field from the leakage current should be lower than $1.4\times 10^{-8}$~T to have the shift on the resonance frequency within 100~Hz. This implies that the leakage current below 100~$\mu$A is acceptable for operation of the OEDM experiment. High-voltage supplies for the electric field $\vec{E}$ can introduce an additional systematic bias to the observable signal. The nonzero voltage drift of the supplies, $dV/dt$, produces a magnetic field induced by the displacement current, flowing in and out of the electrodes, through the relation of $I=CdV/dt$ where $C$ is the capacitance between the vapor cell and electrodes~\cite{Kim:2015}. In case that the voltage drift is very slow, i.e., the first-order drift, it generates a dc magnetic field. This will only affect the shift of the AM resonance frequency, like the leakage current. The oscillating voltage drift at the target frequency will originate a significant systematic in OEDM experiments. Such ripple voltages of high-voltage supplies can be suppressed by, for example, adding a voltage regulator or selecting low-distortion supplies. 

Possible sources of noise are expected to be very small and can be controlled; the ambient high or low frequency noise can be suppressed by using radio-frequency (RF), ferrite, or $\mu$-metal shields.

\section{Sensitivity Estimate}
For the AM-based electron OEDM search, we propose to use a spin-exchange relaxation-free (SERF) AM containing Rb atoms for low frequencies below $\sim$1~kHz and a RF AM containing K atoms for high frequencies above $\sim$1~kHz~\cite{Savukov:2005} for a wide frequency range between 1~Hz to 1~MHz. The AMs are available at Los Alamos National Laboratory. The SERF AM can reach a 1~fT/$\sqrt{\text{Hz}}$ field sensitivity by operating in a near-zero magnetic field and by sufficiently heating the Rb vapor cell to achieve the SERF regime~\cite{Happer:1973,Allred:2002}. Similar high sensitivity of the RF AM is achieved by pumping spins to the stretched state reducing the spin-exchange relaxation (light-narrowing effect)~\cite{Savukov:2005}. Cs magnetometers can also be employed for this electron OEDM search due to the $\sim$ 4 times larger EDM enhancement factor for Cs atoms than Rb atoms~~\cite{nataraj:2008} (discussed below). However, because Cs magnetometers achieved a sensitivity of 40~ $\text{fT}/\sqrt{\text{Hz}}$~[64], roughly 40 times worse than the Rb SERF AM, we do not consider Cs magnetometers here.

\begin{figure}[t]
\centering
\includegraphics[width=0.5\textwidth]{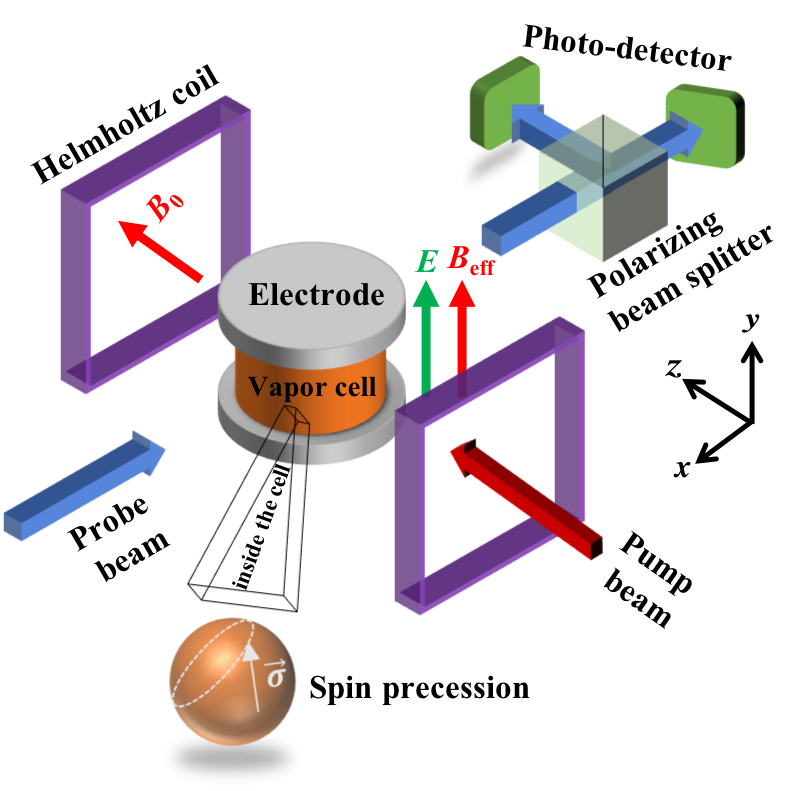}
\caption{Simplified schematic drawing of the axion-induced electron OEDM search using an AM-based experimental method (not scaled). An electric and magnetic fields will be applied to an AM vapor by two electrodes and a Helmholtz coil, respectively. The axion-induced spin precession will be measured with a probe-laser-beam polarimeter, containing a photo-detector.}
\label{fig:setup}
\end{figure}
The schematic representation of a proposed experimental setup is shown in Fig.~\ref{fig:setup}. Two laser beams overlap in an AM vapor cell: a pump beam orients electron spins along its direction and a probe beam reads out the state of the spins. The angle between the pump and probe beam can be arbitrary, e.g., parallel~\cite{Karaulanov:2016} or orthogonal. The conventional configuration of orthogonal beams shown in Fig.~\ref{fig:setup} provides the best field sensitivity. The externally applied holding magnetic field $\vec{B}_0$ is along the pump beam direction, and the external electric field $\vec{E}$ is along the direction perpendicular to both $\vec{B}_0$ and the probe beam. The axion-induced electron spin precession is sensitively detected with a probe-laser-beam polarimeter, containing a polarizing beam splitter and two photo-detectors, as illustraged in Fig.~\ref{fig:setup}, or two polarizers at small angle and a photo-detector. The AM system is placed into magnetic shields made of mu-metal or ferrite to reduce ambient magnetic noise.  

For the Rb and K atoms in the AMs, the effective oscillating magnetic field $\vec{B}_{\text{eff}}$ becomes 
\begin{align}
\gamma\vec{B}_{\text{eff}}(t)=\vec{E}Rd_e (t)
\end{align}
where $R$ is the EDM enhancement factor for the atoms. Similar to the static EDM measurements of atoms and molecules, the OEDM can be also enhanced by the corresponding Schiff moment~\cite{Sandars:1966} for several orders of magnitude~\cite{stadnik:2014}. The gyromagnetic ratio of both Rb and K atoms is $7\times10^{9}$~Hz/T which corresponds to the energy sensitivity of $2.9\times 10^{-20}$~eV based on the AM field sensitivity of 1~fT in an one second measurement. We estimate the sensitivity of our proposed experiment to the electron OEDM. Based on the energy sensitivity of the AMs and Eq.~8, the sensitivity of the energy shift due to the electron OEDM is limited by
\begin{align}
E R_{\text{K/Rb}} d_e=2.9\times 10^{-20}~\text{eV}
\end{align}
where $R_{\text{K/Rb}}$ is the EDM enhancement factor for K and Rb atoms, respectively. Using Eq.~6 and with $E=5$~kV/cm, the sensitivity to the axion-photon coupling, $g_{a\gamma\gamma}$,  becomes
\begin{align}
g_{a\gamma\gamma}=0.07 \Big(\frac{1}{R_{\text{K/Rb}}}\Big) \Big(\frac{m_a}{\text{eV}}\Big) \text{GeV}^{-1}.
\label{eq:EneSen}
\end{align}
For the K AM with $R_{\text{K}}=2.5$~\cite{Sandars:1966}, the sensitivity to $g_{a\gamma\gamma}$ is given by
\begin{align}
 g_{a\gamma\gamma} =2.8\times10^{-2}\Big(\frac{m_a}{\text{eV}}\Big)\text{GeV}^{-1},\label{eq:grrSen1}
\end{align}
while for the Rb AM with $R_{\text{Rb}}=25$~\cite{nataraj:2008},
\begin{align}
 g_{a\gamma\gamma} =2.8\times10^{-3}\Big(\frac{m_a}{\text{eV}}\Big)\text{GeV}^{-1}.\label{eq:grrSen2}
\end{align}

\begin{figure}[t]
\includegraphics[width=0.5\textwidth]{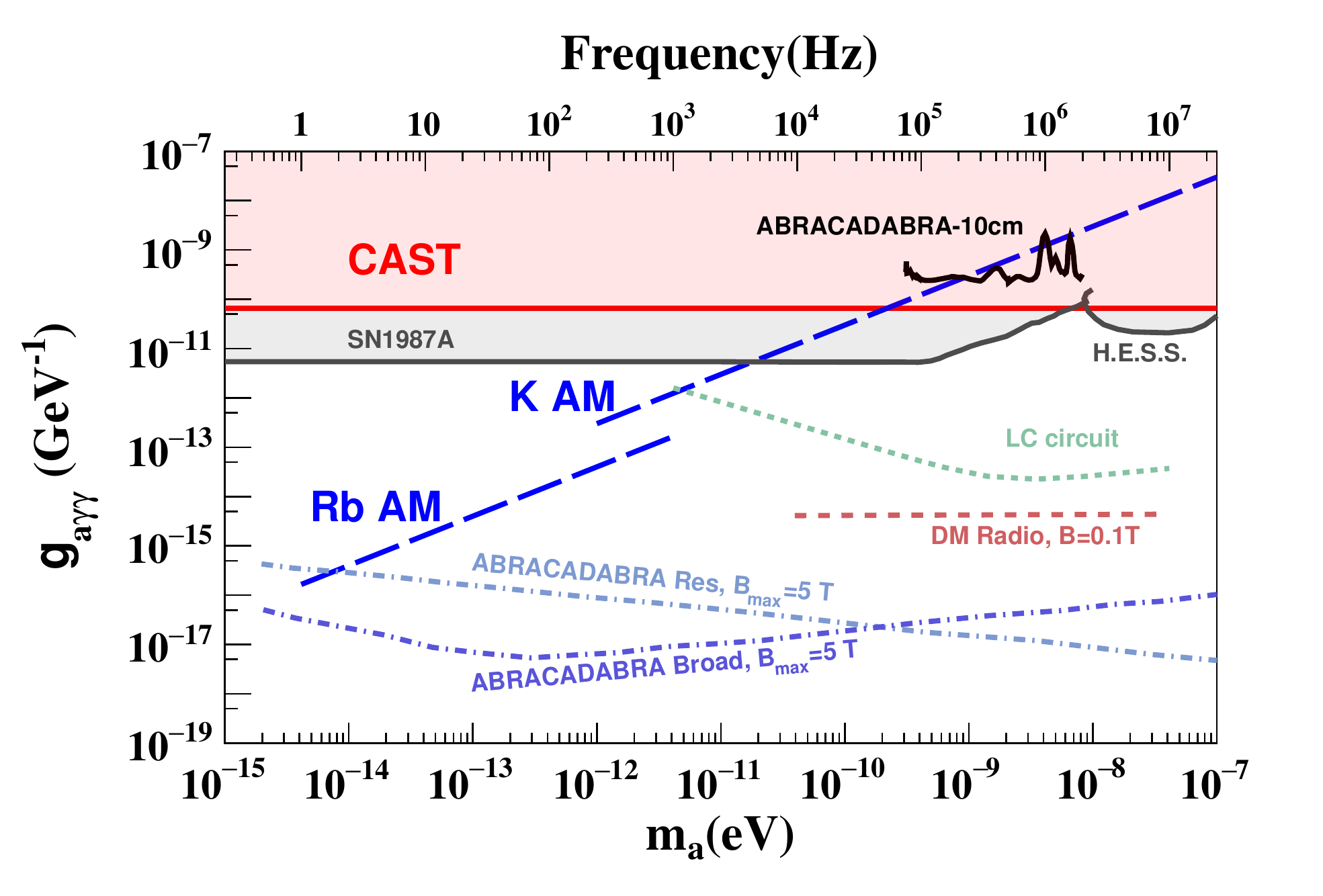}
\caption{Estimated sensitivity of our electron OEDM experiment based on Rb and K AMs to $g_{a\gamma\gamma}$ on the axion mass range with 1 s integration time (blue dashed line). The experiment will improve the current limit set by the CAST experiment~\cite{Anastassopoulos:2017} (red region) for the axion masses between $10^{-14}$ and $10^{-10}$
~eV, specifically by 5 orders of magnitude at $m_a\sim10^{-14}$~eV. The gray region is constrained by astrophysical sources such as High Energy Stereoscopic System (H.E.S.S.)~\cite{Abramowski:2013} and supernova SN 1987A~\cite{Payez:2015}. The sensitivities of other proposed experiments such as ABRACADABRA~\cite{Kahn:2016} with 1~yr integration time, DM Radio~\cite{Chaudhuri:2015,Chaudhuri:2018} with 1.5~yr integration time and LC-circuit~\cite{Chu:2018} with 7~h integration time are also included. The black curve above the CAST limit shows the first result from the experiment of ABRACADABRA-10 cm~\cite{Ouellet:2018}.}
\label{fig:gagg}
\end{figure}

Figure~\ref{fig:gagg} shows the projected sensitivity of the proposed experiment to $g_{a\gamma\gamma}$, based on Eqs.~\ref{eq:grrSen1} and ~\ref{eq:grrSen2}. The proposed experiment can set a new experimental limit on a significant axion mass range between $10^{-15}$ to $10^{-10}$~eV. In particular, the experiments will improve the current experimental bound set by the CERN Axion Solar Telescope (CAST) experiment~\cite{Anastassopoulos:2017} by 5 orders of magnitude at $m_a\sim10^{-14}$~eV. The SERF Rb AM is sensitive at low frequencies from 1 to 1500~Hz~\cite{Savukov:2017}, corresponding to the axion masses from $4.13\times 10^{-15}$ to $6.20\times 10^{-12}~\text{eV}$. This requires the $B_0$ of $2\times 10^{-10}$ to $2\times 10^{-7}~\text{T}$ to match the spin's Larmor frequency to the axion mass.  The upper bound of axion mass range $10^{-10}$~eV, corresponding to the frequency of $\sim$ 10~kHz, requires $B_0$ $\sim3\times10^{-6}$~T.  The sensitivity in the experiment can be significantly improved by orders of magnitude with a long data integration time.

\section{Conclusion}
In conclusion, we proposed a novel, original experimental method to search for the axion-induced oscillating EDM of electrons using very sensitive atomic magnetometers. We estimated that the experiment is sensitive to the axion-photon coupling in the axion masses from  $10^{-15}$ to $10^{-10}$~eV. The electron OEDM experiment can improve the current experimental limit by a few orders of magnitude and push the sensitivity closer to the QCD axion. This approach can be applied to search for OEDMs of other particles such as xenon~\cite{graham:2013}, neutron~\cite{abel:2017}, and molecules~\cite{stadnik:2014} with a non-zero magnetic dipole moment. The method of co-magnetometer readout~\cite{Chupp:1998,Tullney:2013} may be useful to improve the sensitivity in some OEDM experiments as well. In the future, improving technologies for static EDM experiments would also benefit to OEDM experiments.

\section*{ACKNOWLEDGMENTS}
The authors thank Dr. Mike Snow and Dr. Steven Clayton for useful discussions. This work was supported by the U.S. DOE through the LANL/LDRD program.
\bibliography{main}

\end{document}